\documentclass[twocolumn,showpacs,preprintnumbers,amsmath,amssymb]{revtex4}

\usepackage{graphicx}
\usepackage{dcolumn}
\usepackage{bm}

\begin{document}


\title{TeV neutrinos from accreting X-ray pulsars}

\author{W\l odek Bednarek}
 \email{bednar@fizwe4.phys.uni.lodz.pl}
\affiliation{%
Department of Astrophysics, University of \L \'od\'z, 90-236 \L \'od\'z, ul. Pomorska 149/153, Poland
}


\date{\today}

\begin{abstract}
Pulsars inside  binary systems can accrete
matter that arrives up to the pulsar surface provided that
its period is long enough. During the accretion process, matter has to be accelerated to the rotational velocity of the pulsar magnetosphere at the distance where the balance between the pressure of matter
and the magnetic field is achieved. At this distance, a very turbulent and magnetized region is formed in which hadrons can be accelerated to relativistic energies. 
These hadrons can interact with the very strong radiation field coming from the hot polar cap on the neutron star surface created by the in-falling matter. We calculate the neutrino event rates produced in an km$^2$ detector that can be expected from accreting millisecond and classical X-ray pulsars at a typical distance within our own Galaxy.
\end{abstract}

\pacs{97.80Jp,97.60Gb,96.40Qr,98.60Ce,98.70Rz,98.20Hn}
\maketitle

\section{Introduction}

A large number of neutron stars (NS) has been discovered as close companions of low mass and high mass stars (so called low mass and high mass X-ray binaries, see~\cite{liu06,liu07}). These binaries are characterized by a strong X-ray emission that is produced as a result of matter accretion from the companion star onto the neutron star. In some cases, the matter arrives up to the NS surface at the magnetic polar cap region which creates a very hot spot. This class of X-ray binaries shows pulsations correlated with the rotational period of the NS (so called X-ray pulsars, see~\cite{kari07}).  X-ray binaries may contain a long period NS as well as a millisecond
NS. The long period X-ray pulsars are usually found in isolated binary systems whereas the millisecond pulsars (MSPs) are mostly contained within globular clusters~\cite{cr05}. According to the standard scenario, the MSPs are old pulsars spinned up by the angular momentum transfered to the NS from the accreting matter. Many of them have been captured by stars
whose density inside the globular cluster is very large. 
On the other hand, long period X-ray pulsars are found inside binary systems in which NS is formed in a supernova explosion of a more massive companion.  

The matter that is falling onto the NS as a result of the overflow through the inner Lagrangian point usually contains a large angular momentum. This matter forms an accretion disk that can be disrupted at the inner radius due to the pressure of the magnetic field of the NS. 
In the case of accretion from the wind of the massive star, the process occurs more spherically due to the isotropization of the accretion flow by the
shock in the massive star wind.
In some binaries the process of accretion ca be additionally stimulated by the re-radiation of stellar surface of the companion star by the X-ray radiation from the NS. Accreting pulsars can suffer spin up and spin down periods depending on the 
conditions inside the binary system and the parameters of the MSP (i.e. the mass loss rate of the companion star, distance between the stars, neutron star velocity, surface magnetic field strength and the MSP rotational period). These basic parameters determine the specific accretion phase of the matter onto the NS, i.e. whether it is in the accretor, the propeller, or the ejector phase~\cite{lip92}. The ejector phase is characteristic for a fast rotating NS whose magnetized wind is able to prevent any accretion below the light cylinder radius. In the propeller phase, matter can enter the inner pulsar magnetosphere. However, it is blocked by the NS fast rotating magnetosphere at a certain distance from the NS surface due to the centrifugal force. 
Then, most of the matter is expelled from the vicinity of the pulsar and the accretion onto the neutron star surface may occur only episodically when the pressure of the accumulated matter overcomes the pressure of the rotating magnetosphere. 

In this paper we are interested in the acceleration of hadrons during the accretion of matter from the companion star onto the magnetized neutron star.
Hadrons interact with the thermal radiation from the NS surface and produce 
neutrinos which might be observable by the large scale neutrino telescopes
(IceCube, KM3NET). Note that production of neutrinos inside the binary systems
has been up to now considered in a few different scenarios. 
For example, hadrons accelerated in the jet can interact 
with the radiation of the accretion disk~\cite{lw01}, the stellar wind~\cite{rom03} and/or the massive star~\cite{bed05} or hadrons accelerated in the pulsar inner magnetosphere or the pulsar wind shock can interact with the matter of the massive star~\cite{gs85} and/or the accretion disk~\cite{an03}. More recently, production of neutrinos has been considered in the context of the massive binary systems 
detected in the TeV $\gamma$-rays, e.g.~\cite{ch06,th07}.

\section{Pulsar inside binary system}

We are the scenario in which  the accretion process onto the NS surface occurs in the accretor phase. In this phase the matter
passes through the inner pulsar magnetosphere onto the surface of the neutron star. It is characteristic for NSs which are relatively slow rotators with a weak surface magnetic field.
The gravitational energy of the accreting matter is released on the neutron star surface, producing a hot spot around the magnetic pole. The amount of released energy is re-emitted as thermal X-ray emission whose the power can be estimated from:
\begin{eqnarray}
L_{\rm X} = G {\dot M}_{\rm acc} M_{\rm NS}/R_{\rm NS}\approx 2\times 10^{36}M_{16}~~~{\rm erg~s}^{-1},
\label{eq1}
\end{eqnarray}
\noindent 
where ${\dot M}_{\rm acc} = 10^{16}M_{\rm 16}$ g s$^{-1}$ is the accretion rate, and $L_{\rm X}$ is the X-ray thermal emission from the polar cap region. 
The radius and the mass of the NS is assumed to be $R_{\rm NS} = 10^6$ cm and $M_{\rm NS} = 1.4M_\odot$.

The distance from the NS surface at which the magnetic field starts to dominate the dynamics of the in-falling matter (the Alfven radius) can be estimated by comparing the magnetic field energy density to the kinetic energy density of the wind,
$B_{\rm A}^2/8\pi = \rho v_{\rm f}^2/2$,
where $B_{\rm A}$ is the magnetic field in the inner neutron star magnetosphere, 
$\rho = {\dot M}_{\rm acc}/(\pi R_{\rm A}^2v_{\rm f})$ is the density of the accreting matter, $v_{\rm f} = (2GM_{\rm NS}/R_{\rm A})^{1/2}$ is the free fall velocity of the accreting matter, $R_{\rm A}$ is the Alfven radius, and G is the gravitational constant. 

By applying Eq. \ref{eq1}, and assuming that the magnetic field in the neutron star magnetosphere is of the dipole type, i.e. $B_{\rm A} = B_{\rm NS} (R_{\rm NS}/R_{\rm A})^3$, we estimate the location of $R_{\rm A}$ with respect to the NS surface:
\begin{eqnarray}
R_{\rm A} = 7.8\times 10^6 B_{\rm 9}^{4/7}M_{\rm 16}^{-2/7}~~~~{\rm cm},
\label{eq5}
\end{eqnarray}
\noindent
where the magnetic field at the neutron star surface is $B_{\rm NS} = 10^{9}B_{9}$ G. 

Then, we can estimate the magnetic field strength at the transition region,
\begin{eqnarray}
B_{\rm A} = 3.3\times 10^6M_{\rm 16}^{6/7}B_{\rm 9}^{-5/7}~~~~{\rm G}.
\label{eq6}
\end{eqnarray}

The observed thermal luminosity in X-rays, re-radiated from the region of the polar cap, can be calculated from $L_{\rm X} = \pi R_{\rm cap}^2\sigma T_{\rm cap}^4$, where $\sigma$ is the Stefan-Boltzmann constant. In fact, the emission from the polar cap region is well 
described by a black body spectrum (see e.g.~\cite{zan00}).
The radius of the polar cap region on the NS surface, onto which the matter falls, and from which thermal X-ray radiation is emitted, can be estimated from (assuming a dipole structure of the magnetic field):
\begin{eqnarray}
R_{\rm cap} = (R_{\rm NS}^3/R_{\rm A})^{1/2} \approx 3.6\times 10^{5}B_{9}^{-2/7}M_{16}^{1/7} ~~~{\rm cm}.
\label{eq2}
\end{eqnarray}
\noindent
Then, the surface temperature of the polar cap has to be,
\begin{eqnarray}
T_{\rm cap} = (L_{\rm X}/\pi R_{\rm cap}^2\sigma)^{1/4}\approx 1.8\times 10^7B_{9}^{1/7}M_{16}^{5/28}~~{\rm K}.
\label{eq3}
\end{eqnarray}

In general, the accretion of matter  onto the NS can occur provided that 
the radius of the transition region (i.e. the Alfven radius $R_{\rm A}$) lies inside the light cylinder radius of the neutron star, i.e $R_{\rm A} < R_{\rm LC} = cP/2\pi$, where $P$ is the rotational period of the neutron star in seconds , and $c$ is the velocity of light. This condition is fulfilled for, 
\begin{eqnarray}
P > P_{\rm p} = 1.7\times 10^{-3}  B_{\rm 9}^{4/7}M_{\rm 16}^{-2/7}.
\label{eq7}
\end{eqnarray}
\noindent
Moreover, the rotational velocity of the magnetosphere at $R_{\rm A}$ has to be larger than the keplerian velocity of the accreting matter. If the rotational velocity, given by 
\begin{eqnarray}
v_{\rm rot} = 2\pi R_{\rm A}/P\approx 4.8\times 10^{7} B_{9}^{4/7} 
M_{16}^{-2/7}/P~~~{\rm cm~s}^{-1},
\label{eq8}
\end{eqnarray}
is larger than the keplerian velocity,
\begin{eqnarray}
v_{\rm kep} = (GM_{\rm NS}/R_{\rm A})^{1/2}\approx  4.8\times 10^9B_{9}^{-2/7} M_{16}^{1/7}~~~{\rm cm~s}^{-1},
\label{eq9}
\end{eqnarray}
then the matter cannot accrete directly onto the NS surface. It is partially accumulated close to the transition region and partially expelled by the centrifugal force. By comparing Eqs.~\ref{eq8} and~\ref{eq9}, we get the limiting rotational period of the NS below which the matter can accrete onto the NS surface: 
\begin{eqnarray}
P > P_{\rm a} = 0.01 B_{9}^{6/7}M_{16}^{-3/7}.
\label{eq10}
\end{eqnarray}
\noindent
NSs with periods within this range, $P_{\rm p}$ and $P_{\rm a}$, can only accrete matter in the propeller scenario.
However, for periods longer than $P_{\rm a}$, the accretion process occurs 
in the accretor phase. This last stage is the most interesting for this paper.

\section{Acceleration of hadrons}

As we discussed above, the transition region contains turbulent magnetized plasma which provides good conditions for the acceleration of particles. Both electrons and hadrons can be accelerated there. Electrons scatter on thermal radiation from the polar cap region 
and produce $\gamma$-rays by the IC $e^\pm$ pair cascade process as discussed recently in\cite{bed09}.  However, also relativistic hadrons can suffer significant energy losses due to radiative processes in collisions with dense thermal radiation from the polar cap. These
processes produce $\gamma$-rays (that undergo similar processes as the accelerated electrons) and additionally neutrinos that escape from the production region without any absorption. The observation of these neutrinos by large scale detectors will provide crucial information on the relative importance of the acceleration of hadrons
and electrons in the strongly magnetized and turbulent plasma during the accretion process onto the NS.

The maximum power available for the acceleration of particles is limited by the extracted energy in the transition region. This energy can be supplied by two mechanisms. In the case of a quasi-spherical accretion from the stellar wind, the matter has to be accelerated
to the velocity of the rotating magnetosphere at $R_{\rm A}$. The rotating NS decelerates, providing energy to the turbulent region. In the case of accretion through the Lagrangian point, the matter has a large angular momentum, which has to be partially lost in the transition region in order to guarantee the accretion process up to the NS surface. The rotational energy of the accreting matter is then supplied partially to the transition region and to the NS. As a result, the NS reaches the angular momentum and accelerates. 
In the first case, the power which has to be transfered from the rotating NS to the accreting matter can be estimated from:
\begin{eqnarray}
L_{\rm w} = {\dot M}_{\rm acc} v_{\rm rot}^2/2 \approx 10^{31}B_{9}^{8/7}M_{16}^{3/7}P^{-2}~~{\rm erg~s}^{-1}.
\label{eq11}
\end{eqnarray}
\noindent
By Using Eq.~\ref{eq10}, we can estimate the maximum power which can be extracted via accretion from the quasi-spherical wind,
\begin{eqnarray}
L_{\rm w}^{\rm max}\approx 10^{35}B_9^{-4/7}M_{16}^{9/7}~~{\rm erg~s}^{-1}.
\label{eq12}
\end{eqnarray}

In the case of accretion through the accretion disk, the matter arrives to the transition region with the Keplerian velocity. 
This region is now closer to the NS surface than estimated above in $R_{\rm A}$ (see Eq.~{\ref{eq5}) by a factor $\chi\sim 0.1-1$ (see \cite{lamb73}). In order to accrete onto the NS surface, the matter from the disk has to be 
slowed down to the rotational velocity of the NS magnetosphere, i.e. from $v_{\rm kep}$ to $v_{\rm rot}$. Then, the maximum available power extracted in the transition region
is
\begin{eqnarray}
L_{\rm d}\approx {{1}\over{2}}{\dot M}_{\rm acc}(v_{\rm kep}^2 - v_{\rm rot}^2) = 
L_{\rm w}^{\rm max}({{1}\over{\chi}} - {{\chi^2P_{\rm a}^2}\over{P}^2})~~{\rm {{erg}\over{s}}}.
\label{eq13}
\end{eqnarray}
\noindent
which gives $L_{\rm d}\approx L_{\rm w}^{\rm max}/\chi$ for $P\gg P_{\rm a}$.
 
In conclusion,  in both considered here accretion scenarios, from the wind and through the accretion disk, the energy is transfered to the turbulent, magnetized plasma at $R_{\rm A}$. However,
in the first case the pulsar loses its rotational energy but in the second case the pulsar can gain rotational energy from the accreting matter and accelerate its rotation.

A part, $\eta$, of the power, $L_{\rm w}$ or $L_{\rm d}$, can be used for the acceleration of hadrons.
Below we estimate the characteristic energies of accelerated hadrons.
The acceleration rate of hadrons with energy $E$ (and Lorentz factor $\gamma_{\rm p}$) can be  parametrized by a simple scaling to the Larmor radius of particles within a medium with magnetic field B,
\begin{eqnarray}
{\dot P}_{\rm acc} = \xi c E/R_{\rm L}
\approx 1.3\times 10^6\xi_{-1}M_{16}^{6/7}B_{9}^{-5/7} ~~~{\rm erg~s}^{-1},
\label{eq14}
\end{eqnarray}
\noindent
where $\xi = 10^{-1}\xi_{-1}$ is the acceleration parameter, $c$ the velocity of light, 
$R_{\rm L} = E/eB_{\rm A}$ the Larmor radius, and $e$ electron charge. 
The maximum energies of the accelerated hadrons are determined by the balance between the acceleration time scale and the time scale which defines the confinement of hadrons inside 
the turbulent region. The lower limit on this last time scale gives the escape time scale of hadrons
from the acceleration site. It can be defined as, 
$\tau_{\rm esc} = R_{\rm A}/v_{\rm f}\approx 1.1\times 10^{-3}B_{9}^{6/7}M_{16}^{-3/7}~~{\rm s}$.
By comparing this time scale with the acceleration time scale, $\tau_{\rm acc} = m_{\rm p}\gamma_{\rm p}/{\dot P}_{\rm acc}$ (where $m_{\rm p}$ is the proton mass), we can estimate the Lorentz factors to which hadrons can be accelerated,
\begin{eqnarray}
\gamma_{\rm p}\approx 10^{6}\xi_{-1}B_{9}^{1/7}M_{16}^{3/7}.
\label{eq15}
\end{eqnarray}
\noindent
This estimate does not overcome the maximum possible Lorentz factor of particles, which can be accelerated within the region with characteristic diameter $R_{\rm A}$
and magnetic field $B_{\rm A}$, defined by the condition that the Larmor radius is smaller than the Alfven radius, $R_{\rm L} < R_{\rm A}$,
$\gamma_{\rm p}^{\rm max}\approx 6\times 10^{6}B_{9}^{-1/7}M_{16}^{4/7}$.

Hadrons with the Lorentz factors estimated in Eq.~\ref{eq15} are above the threshold
for pion production in collisions with thermal photons from
the polar cap region equal to 
$\gamma^{\rm th}_{\rm p\gamma}\approx 1.6\times 10^4B_9^{-1/7}M_{16}^{-5/28}$.

We can estimate the energy losses of hadrons on collisions with thermal photons from the NS surface inside the transition region,
\begin{eqnarray}
{\dot P}_{\rm p\gamma\rightarrow\pi}  =  \sigma_{\rm p\gamma}cnK E 
\approx  1.5\times 10^6 B_9^{4/7}M_{16}^{33/28} E~~{\rm {{erg}\over{s}}},
\label{eq15b}
\end{eqnarray}
\noindent
where $n = n_{\rm bb}(R_{\rm cap}/((R_{\rm A} - R_{\rm NS}) + R_{\rm cap}))^2$
($n\approx n_{\rm bb}R_{\rm cap}^2/R_{\rm A}^2$ for $R_{\rm A} \gg R_{\rm cap}$) is the density of the black body photons coming from the polar cap at the distance of the transition region $R_{\rm A}$, and $\sigma_{p\gamma}\approx 3\times 10^{-28}$ cm$^{-2}$ and $K\approx 0.15-0.35$ are the cross section and the in-elasticity for pion production due to collision of relativistic protons with thermal photons.
In our estimations we apply the average value for $K = 0.25$. 
As above, we can estimate the maximum possible energies of accelerated hadrons 
due to their energy losses on pion production in collisions with thermal photons
by comparing Eqs.~(13) and (15). We obtain the limit on the Lorentz factor of hadrons, 
\begin{eqnarray}
\gamma_{\rm p\gamma\rightarrow\pi}\approx 1.5\times 10^{6}\xi_{-1}B_{9}^{4/7}M_{16}^{33/28}.
\label{eq15c}
\end{eqnarray}
\noindent
These maximum Lorentz factors of hadrons are typically above the maximum Lorentz factors of hadrons estimated above based on their escape from the acceleration region especially in the case of large accretion rates (note the dependence on the accretion rate in 
Eqs. (14) and (16). Therefore, we conclude that hadrons are at first accelerated 
to maximum energies allowed by the escape mechanism and after that they interact
with thermal photons during their fall onto the NS surface. Note that, density of thermal photons increases towards the NS surface with the distance as $\propto R^2$. Therefore, hadrons interact not very far from the acceleration region.

We consider hadrons which are accelerated with a power law or a mono-energetic spectra at the transition region. In the first case hadrons with energies below the threshold for pion production in collisions with thermal photons also exist. These hadrons can only interact with the matter inside the transition region and the matter accreting onto the NS surface.
We estimate the density of matter at the transition region, $\rho = {\dot M}_{\rm acc}/(\pi R_{\rm A}^2v_{\rm f})\approx 2.2\times 10^{15}B_9^{-6/7}M_{16}^{10/7}~~~{\rm cm}^{-3}$,
and the interaction rate of hadrons for pion production $N_{\rm pp\rightarrow\pi} = \sigma_{\rm pp}c\rho (R_{\rm A}/v_{\rm f})\approx 2\times 10^{-3}M_{16}$.
Therefore, hadron-hadron interactions are not efficient at the transition region.
These lower energy hadrons are convected with the in-falling matter on the NS surface. They interact with much denser matter in the accretion column over the polar cap region. 
However, neutrinos produced in hadron-hadron collisions have on average lower energies due to large pion multiplicity
(see e.g.~\cite{mult}).  We show below that these pions (and muons from their decay), appearing in a very strong magnetic field on the NS surface, can lose energy efficiently via the synchrotron process. Therefore, neutrinos from their decay have too low energies to produce any observable signal in large scale neutrino telescopes.

\section{Production of neutrinos}

We calculate the spectra of muon neutrinos coming from the decay of pions produced in collisions of hadrons with the thermal radiation form the NS surface by applying a numerical code developed for the interaction of hadrons with photons within supernova envelope where the condition are quite similar (see~\cite{bb02}). Two types of hadron injection spectra are considered: (a) a power law differential spectrum with the spectral index 2.1 which cuts at energies estimated by Eq.~\ref{eq15}, and (b) a mono-energetic spectrum with energies given by Eq.~\ref{eq15}. {\bf The second case is considered as a limiting case of the acceleration process of hadrons occurring in the reconnection of magnetic field inside the transition region.
The electric field induced in the reconnection region can in principle accelerate particles to a very
flat spectrum (spectral index lower than 2) in which most of the energy is accumulated with particles at the highest possible energies.}

Pions and muons (from their decay) are produced in a relatively strong magnetic field at $R_{\rm A}$. Therefore, they can 
suffer significant synchrotron energy losses. In order to check whether or not the synchrotron process can change the energies of the produced charged pions and muons before their decay, 
we calculate their synchrotron energy loss time scales and compare them with their life times.
From this comparison, we put an upper limit on the Lorentz factor of pions, 
\begin{eqnarray}
\gamma_{\pi}^{\rm s}\approx 7\times 10^7 B_{\rm 9}^{10/7}M_{\rm 16}^{-12/7},
\label{eq15c}
\end{eqnarray}
\noindent
that are able to decay before losing a significant amount of energy.
We take the effects of synchrotron energy losses of pions on the produced spectra of neutrinos in the case of their production by hadrons  with the Lorentz factors above $\gamma_\pi^{\rm s}$ into account by simply replacing 
their $\gamma_\pi$ by $\gamma_\pi^{\rm s}$.
However, in the case of muons, the critical value of the Lorentz factors is about two orders of magnitude lower. Therefore, neutrinos from muons decays 
have rather low energies. They are not included in our calculations of the neutrino event rates.

For the example calculations, we fix the following parameters describing acceleration of hadrons: $\xi_{-1} = 0.1$, and $\eta = 0.1$. In table~1, we show the muon neutrino event rates that are expected in a km$^{2}$ neutrino detector from millisecond pulsars at the distance of 5 kpc which accrete matter from the stellar wind. The typical parameters for the MSPs and the accretion rates are considered.
The results are shown for a power law spectrum of hadrons (marked by {\it power})  and a mono-energetic spectrum (marked by {\it mono}).
The number of neutrino events is estimated by integrating the muon neutrino spectra over the probability of their detection,
\begin{eqnarray}
N_{\mu} = {{S}\over{4\pi D^2}} \int_{E_\nu^\textrm{min}}^{E_\nu^\textrm{max}} 
P_{\nu\rightarrow\mu}(E_\nu) {{dN_\nu}\over{dE_\nu dt}} dE_\nu,
\label{eq20}
\end{eqnarray}
where $S = 1$ km$^2$ is the surface of the detector, $D$ is the distance to the source, $P_{\nu\rightarrow\mu}(E_\nu)$ is the energy dependent detection 
probability of a muon neutrino \cite{prob}, $dN_\nu/dE_\nu dt$ is the differential neutrino spectrum produced at the source,
$E_\nu^\textrm{min} = 1$ TeV is the minimum energy of produced neutrinos 
over which we integrate the event rate in the 1 km$^2$ neutrino telescope and the maximum energy of neutrinos produced in this model is typically in the range $E_\nu^\textrm{max} = 10^4 - 10^5$ GeV. This maximum energy of neutrinos can not be simply expressed since it depends on the parameters of the model (see Eqs.~(14) and (17)). Therefore, it is obtained numerically in our calculations.
The expected neutrino event rates from millisecond pulsars is up to a few per km$^2$ per yr (depending on the model). These event rates should be detected significantly by the IceCube neutrino detector. 

\begin{table}[t]
  \caption{Neutrino event rates from a MSP at 5 kpc distance.}
  \begin{tabular}{llllll}
\hline 
\hline 
\\
 B [G]~~~ &  M [g s$^{-1}$]~~ &  $\gamma_{\rm p}^{\rm max}$~~~  & ~P$_{\rm a}$ [ms] & ~~~N$_{\nu}$ [km$^{-2}$yr$^{-1}$]~~~~  \\
              &          &      &      &  ~~~power~~~mono  \\
\hline
\\
 $3\times 10^{8}$   &  $10^{17}$            &   $2.3\times 10^5$    &   ~~~~13     &  ~~~1.0~~~~~~~5.1 \\
\\
 $10^9$             &  $3\times 10^{17}$    &   $4.3\times 10^5$   &    ~~~~23     & ~~~2.1~~~~~~~5.6 \\
\\
 $3\times 10^{9}$   &  $10^{18}$            &   $8.4\times 10^5$    &   ~~~~36     & ~
~5.8~~~~~~~~5 \\
\\
\hline
\hline 
\end{tabular}
  \label{tab1}
\end{table}

We also show the neutrino event rates that are expected from the two nearby X-ray pulsars inside the binary systems GROJ 1744-28 and A 0535+262, which are characterized by long rotational periods, strong surface magnetic fields, and high accretion rates. Also from these sources a few neutrino events per year are expected in an IceCube size neutrino detector.

\begin{table}[t]
  \caption{Neutrino event rates from X-ray pulsars.}
  \begin{tabular}{lllll}
\hline 
\hline 
\\
Name   &  B [G]  &  $L_x$ [erg s$^{-1}$] &  D [kpc]    & N$_{\nu}$ [km$^{-2}$yr$^{-1}$] \\
       &           &                         &             &  power~~mono \\
\hline
\\
GRO 1744-28 &   $2\times 10^{11}$  &  $2\times 10^{38}$  &  4    & 1.6~~~~~2.8 \\
\\
A 0535+262   & $4\times 10^{12}$  &  $9\times 10^{37}$  & 1.8   & 0.4~~~~~0.8  \\
\\
\hline
\hline 
\end{tabular}
  \label{tab2}
\end{table}

%
%
\section{Discussion and Conclusion}

We have considered the model in which hadrons are accelerated within the inner
pulsar magnetosphere as a result of the interaction of accreting matter with the rotating magnetosphere.
The acceleration process occurs in the turbulent transition region. Hadrons produce neutrinos mainly in collisions with thermal radiation arriving from the hot spot on the NS surface. These hadrons can not escape from the NS vicinity since they accrete onto the NS surface. Therefore, hadrons accelerated in such a model can not contribute to the cosmic ray content within the Galaxy.
The production process of neutrinos is accompanied by the production of $\gamma$-ray photons with similar spectra and energies. However, these $\gamma$-rays initiate an Inverse Compton $e^\pm$ pair cascade in the strong thermal radiation from the NS surface (as considered recently by Bednarek~\cite{bed09}). Only $\gamma$-rays with MeV-GeV energies can  
escape from the radiation field of the hot NS surface. Therefore, we propose that the neutron stars accreting at high rates can be observable by the recently launched {\it Fermi} LAT detector in the GeV $\gamma$-rays but also by the large scale neutrino detectors such as IceCube and KM3NET.

As an example we estimated the neutrino event rates in the km$^2$ detector
for two relatively nearby binary systems which accrete at large rate, GRO 1744-2
and A 0535+262. Other considered sources are expected to emit lower neutrino fluxes
due to the larger distances or lower accretion rates. We also estimate the neutrino event rates expected from the millisecond pulsars at the typical distance of globular clusters. It is clear (from Table 1) that some accreting millisecond pulsars could become detectable neutrino sources.
 
Different factors can enhance or reduce the neutrino event rates reported in Tables~1 and~2. The obvious ones are related to the distance to the source and the variable 
neutrino emission due to a change in the accretion rate caused e.g. by the eccentric orbit of the neutron star. Moreover, in the case of the accretion process through the accretion disk, the power available for the acceleration of particles can be enhanced by a factor of $\chi^{-1}$ ($\chi$ is typically in the range $\sim 0.1-1$, see Eq.~12) with respect to the case of the quasi-spherical accretion.  Therefore, the neutrino event rates reported in Table 1 should be scaled as well by that factor. From another site, most of the observed millisecond pulsars belong to globular clusters. It is expected that a massive globular cluster can contain up to 
$\sim 100$ MSPs\cite{tav93}. In fact, inside the globular clusters Tuc 47 and Ter 5, already 
$\sim 20-30$ MSP have been discovered~\cite{cr05}. A significant amount of the MSP cannot be discovered directly by their pulsed radio emission since they are inside compact binary systems (so called hidden MSPs\cite{tav91}). These hidden MSP can become neutrino sources via the production mechanisms,
that have being discussed in our model.
Therefore, the neutrino event rates expected from the whole population of MSPs in a specific globular cluster can be enhanced due to the cumulative contribution from many sources. On the other hand, non-observation of neutrinos from the globular clusters give important constraint on the number of hidden MSPs in these objects.

\begin{acknowledgments}
This work is supported by the Polish MNiSzW grant N N203 390834. 
\end{acknowledgments}

\end{document}